\documentclass[journal,twocolumn]{IEEEtran}

\usepackage{graphicx}
\usepackage{psfrag}
\usepackage{amsmath}
\usepackage{amsfonts}
\usepackage{amssymb}
\usepackage{optidef}
\usepackage{makecell}
\usepackage{algpseudocode,algorithm,algorithmicx}
\usepackage{kotex}
\usepackage{hyperref}

\usepackage{empheq}
\usepackage{optidef}
\usepackage{bm}
\usepackage{algorithm}
\usepackage{algpseudocode}
\usepackage{multirow}

\usepackage[table,xcdraw]{xcolor}
\usepackage{tcolorbox}
\usepackage{balance}

\begin{document}

\title{
Learning-based Ecological Adaptive Cruise Control of Autonomous Electric Vehicles: A Comparison of ADP, DQN and DDPG Approaches
}
\author{{Sunwoo Kim} and {Kwang-Ki~K.~Kim}
	\thanks{Manuscript received xx xx, 2022; revised xx xx, 2022.}        
	\thanks{The authors are with the Department of Electrical and Computer Engineering, Inha University, Incheon 22212, Republic of Korea.}
	\thanks{Corresponding author: K.-K.~K. Kim ({\tt kwangki.kim@inha.ac.kr})}
	\thanks{\small Corresponding author: 김광기 ({\tt kwangki.kim@inha.ac.kr}) This research was partly supported by the National Research Foundation of Korea (NRF) grant (NRF-2020R1F1A1076404) and Korea Institute for Advancement of Technology (KIAT) grant funded by the Korea Government (MOTIE) (P0017124, The Competency Development Program for Industry Specialist).}
}


\maketitle

\begin{abstract}
This paper presents model-based and model-free learning methods for economic and ecological adaptive cruise control (Eco-ACC) of connected and autonomous electric vehicles. For model-based optimal control of Eco-ACC, we considered longitudinal vehicle dynamics and a quasi-steady-state powertrain model including the physical limits of a commercial electric vehicle. We used adaptive dynamic programming (ADP), in which the value function was trained using data obtained from IPG CarMaker simulations. For real-time implementation, forward multi-step look-ahead prediction and optimization were executed in a receding horizon scheme to maximize the energy efficiency of the electric machine while avoiding rear-end collisions and satisfying the powertrain, speed, and distance-gap constraints. For model-free optimal control of Eco-ACC, we applied two reinforcement learning methods, Deep Q-Network (DQN) and Deep Deterministic Policy Gradient (DDPG), in which deep neural networks were trained in IPG CarMaker simulations. For performance demonstrations, the HWFET, US06, and WLTP Class 3b driving cycles were used to simulate the front vehicle, and the energy consumptions of the host vehicle and front vehicle were compared. In high-fidelity IPG CarMaker simulations, the proposed learning-based Eco-ACC methods demonstrated approximately 3--5\% and 10--14\% efficiency improvements in highway and city-highway driving scenarios, respectively, compared with the front vehicle. A video of the CarMaker simulation is available at \href{https://youtu.be/DIXzJxMVig8}{https://youtu.be/DIXzJxMVig8}.
\end{abstract}

\begin{IEEEkeywords}
Eco-driving,
Fuel efficiency,
Electric vehicles, 
Optimal control, 
Dynamic programming, 
Approximate dynamic programming,
Reinforcement learning
Function approximation.
\end{IEEEkeywords}
	
\IEEEpeerreviewmaketitle

\section{Introduction}
\label{sec:1}
With the recent increase in interest in environmental pollution, many countries and organizations are regulating vehicle emissions. Accordingly, the proportion of vehicles powered by electric drive systems is gradually increasing~\cite{Li2019review}. In addition to battery electric vehicles (BEVs), hybrid electric vehicles (HEVs), plug-in hybrid electric vehicles (PHEVs), and fuel-cell electric vehicles (FCEVs) are other types of electrified vehicles (xEVs)~\cite{Chan2007}. Despite technical advances, xEV range anxiety remains a major obstacle that must be overcome.

For xEV range extension, there are two main research directions: fuel-economic speed planning, which is also known as ecological or economic driving (eco-driving), and fuel-economic routing (eco-routing)~\cite{watzenig2017comprehensive}. Various research and development efforts have been made to improve fuel economy~\cite{Guzzella2013,Vahidi2018,Boulanger2011,Ehsani2018}. Because battery electric vehicles take longer to charge than conventional vehicles, increasing the travel distance per charge by improving fuel economy is one of the most important research issues~\cite{Sciarretta2015,guanetti2018control}. There are several approaches to obtain an energy-efficient acceleration/deceleration strategy, taking into account the constraints on the powertrain: dynamic programming (DP)~\cite{Ozatay2014,kim:access:2021,kim:tits:2021}, Pontryagin's maximum principle (PMP)~\cite{Shen2020PMP}, model predictive control (MPC)~\cite{Lim2017} and reinforcement learning (RL)~\cite{Lee2020RL}.

The recent development of communication technologies has encouraged vehicle-to-everything (V2X) applications.
With the development of connectivity and autonomous vehicle technologies, the expectation that higher energy efficiency will be obtained through eco-driving is increasing. However, it is difficult to achieve the theoretical maximum energy efficiency for all hours of driving, unless the vehicle is driving alone, or without any constraints such as a road speed limit or forward vehicle. In particular, in countries with many vehicles, such as South Korea, the acceleration/deceleration strategy is limited not only on driving downtown, but also on highways. Therefore, it is necessary to develop an energy-optimal acceleration/deceleration strategy that considers proceeding vehicles, called ecological adaptive cruise control (Eco-ACC) technology.

For the optimal control problem (OCP) associated with Eco-ACC, the objective is to determine optimal acceleration and deceleration strategies that minimize energy consumption while satisfying collision avoidance. Many classical optimal control methods have been applied to the solution methods of the Eco-ACC OCP, for example, PMP~\cite{watzenig2017comprehensive}, DP~\cite{kim:tits:2021,Bae2019DP}, and MPC~\cite{Jia2020MPC}. DP has a property that its solution is globally optimal, and its associated value function can be obtained in a sequential manner~\cite{Bertsekas2005}, which follows the principle of optimality~\cite{Bellman1954}. Additionally, an optimal solution can be obtained with nonlinear constraints for all feasible states using backward induction. However, the computation in DP is inefficient, and the complexity increases exponentially as the state and input dimensions increase, which is called the curse of dimensionality. Such computational requirements render the DP solution infeasible for real-time planning and control.

To resolve the real-time implementation issues of the DP framework, this study considers a method of approximate dynamic programming (ADP) in which the unknown value function is approximated by a universal function~\cite{Bertsekas2005}. There are many different types of universal functions for the approximation~\cite{Powell2007}. In~\cite{JOHANNESSON2008}, the value function was approximated using a piecewise linear function and model approximation. In~\cite{Deisenroth2008,Deisenroth2009,bae2022gaussian}, a non-parametric function approximation method using Gaussian process regression (GPR) was applied to approximate the value function for adaptive optimal control. As many recent algorithmic advances in reinforcement learning (RL) have been made, attempts have been made to apply RL techniques to eco-driving. In~\cite{Zhou2020DDPG}, deep deterministic policy gradient (DDPG), a method that combines the advantages of the value-based method with those of the policy-based method, was applied to provide an idea for eco-driving in paths that include signalized intersections, and twin delayed DDPG (TD3) has been applied for eco-driving in an urban driving scenario~\cite{Wegener2021TD3}.

In this study, an acceleration and deceleration strategy for Eco-ACC of autonomous electric vehicles was established based on the reinforcement learning method, and ideas for improving the energy efficiency are presented. Therefore, an optimal control problem for the purpose of eco-ACC was defined, and the speed profile as an optimal solution was obtained through (i) DP, (ii) ADP, and (iii) RL for various conditions. A simulation environment was established to measure the travel distance and energy efficiency based on the speed profile obtained using the three different methods. The Eco-ACC strategy can be discovered for electric vehicles based on the simulation results.

The remainder of this paper is organized as follows: Section~\ref{sec:vehmodel} provides the mathematical models of vehicle dynamics and the EV powertrain. In Section~\ref{sec:ocp}, we formulate optimal control problems for maximizing the fuel efficiency of the vehicle while following the front vehicle. Section~\ref{sec:backgrounds} briefly introduces basic solution methods customized to solve the OCP problems of Eco-ACC. Section~\ref{sec:method} describes the proposed Eco-ACC strategies and their implementation procedures. The CarMaker simulation results of applying the proposed Eco-ACC strategies are analyzed by performing comparisons in Section~\ref{sec:sim}. Finally, concluding remarks are provided in Section~\ref{sec:conclusion}.

\section{Vehicle Modeling}
\label{sec:vehmodel}

\begin{figure}[t]
	\centering
	\includegraphics[width=.9\linewidth]{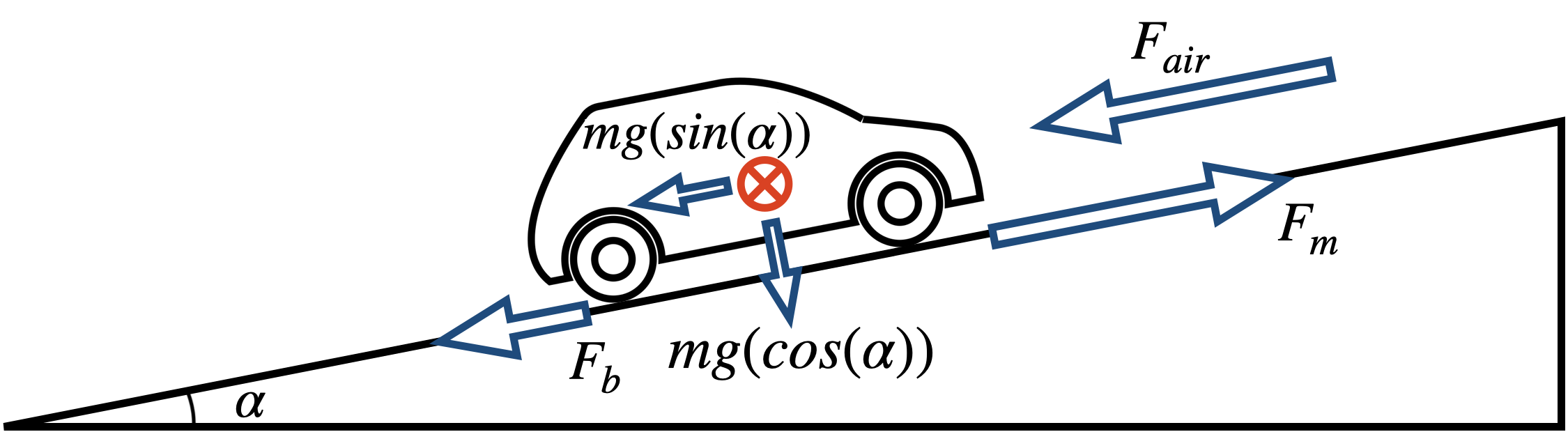}\vspace{-1mm}
	\caption{Simple longitudinal dynamics of vehicle.}
	\label{fig:EV_fig}
\end{figure}
\begin{figure}[t!]
	\centering
	\includegraphics[width=.92\linewidth]{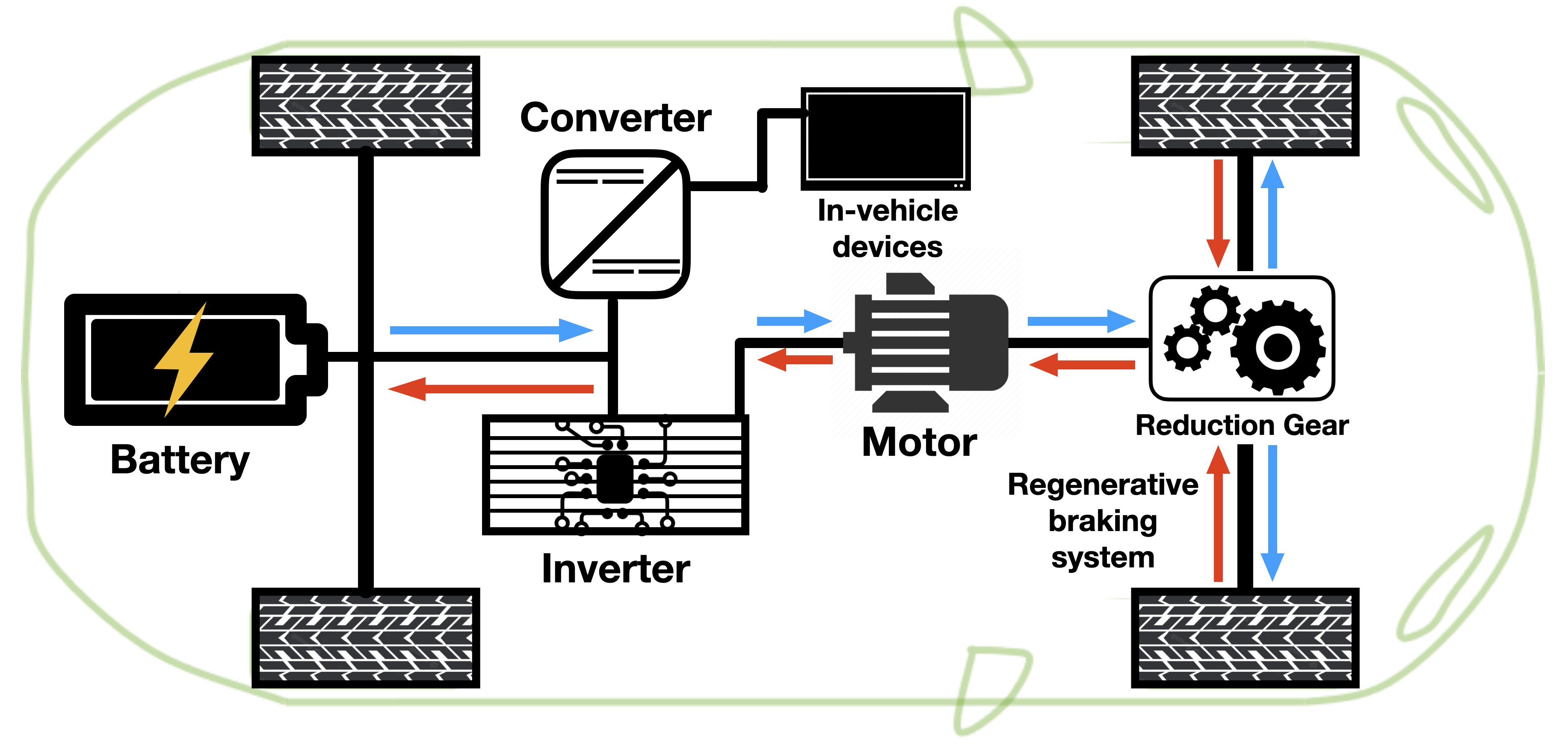}\vspace{-2mm}
	\caption{Simple EV powertrain and drivetrain architecture.}
	\label{fig:simple_powertrain}
\end{figure}
\begin{figure}[t!]
	\centering
	\includegraphics[width=.825\linewidth]{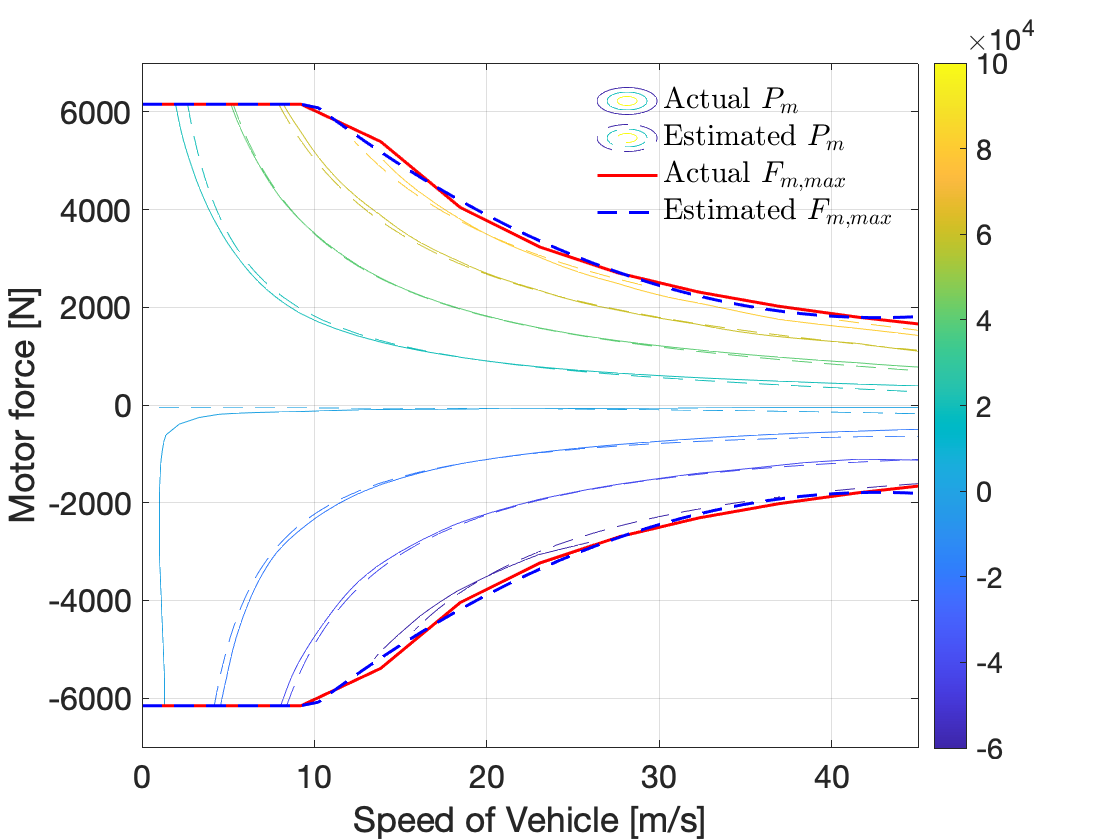}\vspace{-2mm}
	\caption{Motor power according to vehicle speed and motor force.}
	\label{fig:motorpower_map}
\end{figure}

\subsection{Vehicle longitudinal dynamics}
For the dynamical system modeling of Eco-ACC, we consider the following longitudinal vehicle dynamics:
\begin{align}
	m\dot{v}(t) = F_m(t)-F_b(t)-F_{air}(t)-F_{\alpha}(t) \,,
\end{align}
where $v(t)$ is the longitudinal speed of the vehicle, and $m$ is the vehicle mass.
Load force $F_{air}(t)$ is the longitudinal aerodynamic drag. Assuming that the wind speed is relatively small, the aerodynamic drag force can be expressed as
\begin{align}
	F_{air}(t) = \frac{1}{2}\rho_a C_d A_f v^2(t) \,,
\end{align}
where $\rho_a$ is the mass density of air, $C_d$ is the aerodynamic drag coefficient, $A_f$ is the frontal area of the vehicle. Force $F_{\alpha}(t)$ refers to the combined road load
\begin{align}
	\displaystyle F_{\alpha}(t) = mg  \left(\sin(\alpha(D(t)))+C_r\cos(\alpha(D(t)))\right) \,,
\end{align}
where $g$ is the gravitational acceleration, $C_r$ is the rolling resistance coefficient, $\alpha(s(t))$ is road slope, and $s(t)$ is the travel distance.

\subsection{EV Powertrain system}
The EV powertrain consists of a battery, inverter, converter, reduction gear and one motor, and regenerative braking system, as shown in in Fig. ~\ref{fig:simple_powertrain}. $P_{\rm mot}$ is the power that the motor receives from the battery or the power that the generator provides to the battery. Compared with the motor traction power, the loss in the power transmission, $P_{\rm loss}$, and the power consumed by other in-vehicle auxiliary devices, $P_{\rm aux}$, are relatively small, and the battery charging/discharging power is simplified as
\begin{equation}\label{eq:powercomplex}
	P_{\rm  batt} = P_{\rm motor}+P_{\rm loss}+P_{\rm aux} \approx P_{\rm mot} \,.
\end{equation}

From the rotational speed of the motor and wheel torque, the relation between the vehicle speed and force acting on the wheel can be obtained as
\begin{align}
	F_m(t) &= \frac{T_m}{r_w}, \quad v(t) = \frac{r_w}{\gamma} w_m(t) \,,
\end{align}
where $\gamma$ and $r_w$ are the final gear-ratio and wheel radius, respectively.

For modeling and simulation, the electric motor data imported from ADVISOR~\cite{ADVISOR} were analyzed to determine the motor's regression model for energy consumption using the least squares method. The regression model was applied to the fifth-order polynomial in terms of $v(t)$ and $F_m(t)$:
\begin{equation}\label{eq:reg}
\begin{split}
	&P_{\rm mot}(v(t),F_m(t))\\
	&\ = (c_1v+c_2v^3+c_3v^5+c_4vF_m+c_5vF_m^2+c_6v^3F_m) .
\end{split}
\end{equation}
In the model fitting, the relative root mean square error (RRMSE) of the regressor~\eqref{eq:reg} was 0.1917, for which the actual experimental data were used for training and validation.

The maximum force that the motor can continuously generate is determined by the rotational speed of the motor, as shown in Fig. ~\ref{fig:motorpower_map}. Without tire slip, the speed of the vehicle and the force acting on the tire have a one-to-one correspondence and a constant value in the low vehicle-speed range, but tend to be inversely proportional to the speed of the vehicle in the high-speed range.
The following regression models were used in this study:
\begin{equation}\label{eq:limf}
\begin{split}
	F_{m, \max}(v)&= \min \{ \overline{F},c_0+{c_1 \over v}\} \,, \\
	F_{m, \min}(v) &= \max \{ \underline{F},-c_0-{c_1\over v}\} \,,
\end{split}
\end{equation}
where $\overline{F}$ and $\underline{F}$ denote the maximum and minimum motor forces in the low-speed range, respectively.
In model fitting, the RRMSE of the regressor~\eqref{eq:limf} was 0.3000, for which the actual experimental data were used for training and validation.

\section{Optimal Control Problem} 
\label{sec:ocp}
For Eco-ACC, we considered the following constrained optimal control problem (OCP):
\begin{align}
	\underset{F_m(\cdot),F_b(\cdot)}{\mbox{minimize}} \quad \quad \quad \quad \,  J &= \int_{0}^{t_f} P_{mot}(v(t),F_{m}(t)) dt \nonumber \\
	\mbox{subject to} \quad \quad 
	{d\over dt} d(t)& = v_f(t)-v(t) \nonumber \\
	m{d\over dt}v(t) &= F_m(t)\!-\!F_b(t)\!-\!F_{air}(t)\!-\!F_{\alpha}(t) \nonumber \\
	v & \in [0, 40] \nonumber \\
	F_{m,\min}(v(t)) &\le F_m(t) \le F_{m,\max}(v(t)) \label{eq:objf} \\
	F_{b,\min} &\le F_b(t) \le F_{b,\max} \nonumber \\
	F_m(t)F_b(t) &\le 0 \nonumber  \\
	d(t) &\in (0, 2000] \nonumber \\
	v(0) &= 0 \nonumber \\
	d(0) &= 50 \nonumber,
\end{align}
where the host vehicle maintains a distance gap $d(t)$ between $0$ m and $2000$ m from the front vehicle and minimizes the energy in this process. This is considered as the objective function to be minimized in the optimal control problem.

\section{Backgrounds in Solution Methods} 
\label{sec:backgrounds}
\subsection{Dynamic Programming}
Dynamic programming was introduced by Bellman~\cite{BellmanDP} as an algorithm that can find the global optimal solution of an optimal control problem. DP is based on the \emph{principle of optimality}, that is, the optimization in future does not depend on the initial conditions and past-control inputs. The algorithm solves complex problems by dividing them into several small subproblems, and it reuses the solution of the subproblems in the process of solving complex problems.

\subsection{Approximate Dynamic Programming}
Approximate dynamic programming was introduced to resolve the problem of high computational resource requirements and the curse of dimensionality. Because the value function obtained in DP is calculated for discrete state space, it cannot be applied to a problem defined in a continuous state space. To solve this problem, ADP uses an approximate value function to use the value function calculated from the discrete state space in the continuous state space. Unlike DP, a suboptimal solution can be obtained using this method. The methods for obtaining the approximate value function are (i) polynomial regression, (ii) Gaussian process regression, and (iii) neural network approximation.

\subsection{Reinforcement Learning}
\begin{figure}[t]
	\centering
	\includegraphics[width=1\linewidth]{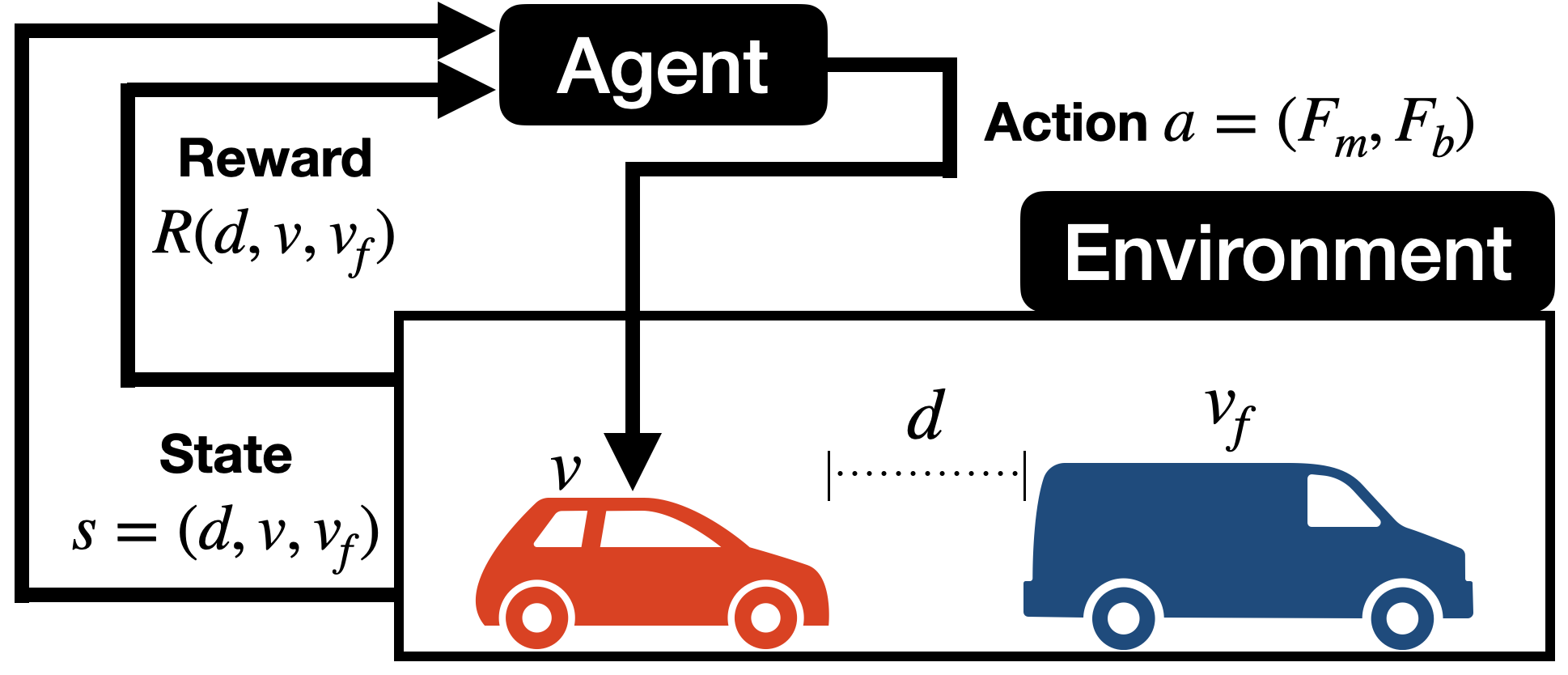}\vspace{-2mm}
	\caption{Basic framework of model-free RL-based Eco-ACC.}
	\label{fig:basic_rlacc}
\end{figure}

In this study, a typical framework of reinforcement learning~\cite{Sutton1998} for Eco-ACC was configured, as shown in Fig. ~\ref{fig:basic_rlacc}. The agent acts to maximize the cumulative rewards from the environment during the entire episode. At time $t$, the agent receives an observation (or state) $s_t$ from the environment, performs action $a(t)$ according to the current policy $\pi$, and receives information about the state $s_{t+1}$ and reward $r_{t+1}$ from the environment; the same procedure is repeated until the end of the episode. Meanwhile, several reinforcement-learning techniques, such as value-based and policy-based RL, are the methods for the agent to learn.

For the RL agent, Deep Q-Network (DQN)~\cite{Mnih2013DQN} and Deep Deterministic Policy Gradient (DDPG)~\cite{Lillicrap2016DDPG} were applied to solve the OCP~\ref{eq:objf} without using any models but only using data. 
Both RL methods are model-free and off-policy algorithms that use the Q-function approximation and the replay buffer.
DQN is presented to solve the problem of unstable learning and divergence due to moving target value in Q-learning by introducing the replay buffer and the target network, but it can only handle discrete and low-dimensional action spaces.
DDPG is an algorithm that combines the advantages of value-based and policy-based RL, which can overcome the limitations of the inability to directly apply Q-learning in continuous action spaces. The main idea of DDPG originates from the Deep Q-Network (DQN) algorithm ~\cite{Mnih2013DQN}; however, DDPG uses two additional ideas. First, the algorithm uses the idea of actor--critic~\cite{Konda2000ac}, that two different neural networks approximated by an actor network and a critic network, each of which has two networks, a main network and a target network, to train more softly. Second, instead of the $\epsilon$-greedy method, DDPG adds noise to the deterministic action to promote exploration. The DDPG learning process is described in Algorithm~\ref{alg:ddpg}.

\begin{algorithm}[t]
	\caption{${\tt DDPG}$}\label{alg:ddpg}
	{Randomly initialize critic $Q(s,a|\theta^Q)$ and actor $\mu(s|\theta^\mu)$}\\
	{Initialize target networks :}\\
	$Q' \gets Q$, $\mu' \gets \mu$\\
	{Initialize replay buffer $R$}
	\begin{algorithmic}
		\For {episode =1, M}\\
		\quad {Initialize a random process $\mathcal{N}$ for action exploration}\\
		\quad {Receive initial state $s_1$}
		\For {t=1, T}\\
		\qquad {Execute $a_t = \mu(s_t|\theta^\mu) + {\mathcal {N}}_t$ and observe $r_t$ and $s_{t+1}$}\\
		\qquad {Store $(s_t,a_t,r_t,s_{t+1})$ in $R$}\\
		\qquad {Sample a random minibatch of $N$ transitions :}\\
		\qquad $(s_i,a_i,r_i,s_{i+1})$ from $R$\\
		\qquad {Set $y_i = r_i + \gamma Q'(s_{i+1},\mu'(s_{i+1}|\theta^{\mu'}) | \theta^{Q'})$}\\
		\qquad {Update critic by minimizing the loss :}\\
		\qquad $L = {1\over N}\sum_{i} (y_i - Q(s_i,a_i|\theta^Q))^2$\\
		\qquad {Update the actor policy using the sampled PG :}\\
		\qquad $\nabla_{\!\theta^\mu} J \approx {1\over N} \sum_{i} \!\nabla_{\!a} Q(s,a|\theta^Q)|_{(s_i,\mu(s_i))} \!\nabla_{\!\theta^\mu} \mu(s|\theta^\mu)|_{s_i}$\\
		\qquad {Update the target networks $Q'$ and $\mu'$}
		\EndFor
		\EndFor
	\end{algorithmic}
\end{algorithm}

\section{Solution Methods} 
\label{sec:method}
To determine the energy-efficient acceleration/deceleration strategy, methods of DP, ADP, and RL were applied to find a solution for the optimal control problem described in Section~\ref{sec:ocp}.

\subsection{Dynamic Programming}
For the DP method, two methods are used to obtain the optimal solution:(i) DP-Backward and (ii) DP-Forward. In DP-Backward, a value table corresponds to the optimal cost-to-go from a set of quantized states. In DP-Forward, multistep look-ahead finite-horizon optimization is performed, in which a terminal-state penalty function approximates the optimal-cost-to-go. 

In this study, the state variables and control inputs are quantized as follows:
\begin{equation}
\begin{split}
	x_k &\in Q_x = \{x^1, \cdots , x^{n_d \times n_v \times n_{v_f}}\} \,, \\
	u_k &\in Q_u = \{u^1, \cdots , u^{n_{F_m} \times n_{F_b}}\} \,, 
\end{split}
\end{equation}
where $x = (d, v, v_f)$ and $u = (F_m, F_b)$ are the state and control vectors, respectively.

\subsubsection{DP-Backward (Off-policy for computing the value table)}
\label{sec:method:DPB}
In the DP-Backward step, the input and state variables are quantized to calculate the value (i.e., optimal cost-to-go) table for $V\!(d, v, v_{f})$
\begin{equation}
	\quad V (d, v, v_{f}) = \underset{{u}\in \bm{U}}{\min} \left\{ C(u,v) + V(d^{+}, v^{+}, v_{f}^{+}) \right \} \,, \\
\end{equation}
where the stage cost is $C(u,v) = P_{motor}(v,F_m) \Delta t$ and $\Delta t$ is the equi-spaced time interval of the sampling instances. 
In the DP-forward process, the speed profile of the host vehicle can be determined using the value table and the profile of the front vehicle. In this study, two speed profiles were generated under different conditions using the DP-based method:
\begin{itemize}
\item
\emph{DP-Forward I}: One-step look ahead policy without considering any acceleration nor deceleration constraints; 
\item
\emph{DP-Forward II}: Two-step look ahead policy with the value table computed offline.
\end{itemize}
They have different prediction horizons and control constraints, and their performances are compared in Section~\ref{sec:sim}.

\subsubsection{DP-Forward I (One-step look ahead)}
\label{sec:method:DPF1}
With the value table obtained from the DP-Backward process, the optimal control input $u^*=\left(F_{m}, F_{b}\right)$ and corresponding speed $v^{+*}$ can be determined as
\begin{align}
	\qquad &u^* = \underset{u \in \bm{U}}{\mathop{\arg\min}} \left\{ C(u,v)+V(d^+,v^+,v_f^+) \right\} \,, \nonumber \\
	&v^{+*} = v + a(u^*) \Delta t  \,.
\end{align}

\subsubsection{DP-Forward II (Two-step look ahead)}\label{sec:method:DPF2}
The third DP-based methods of car-following speed planning is the two-step look ahead policy with the value table computed offline.
By predicting the speed of the front vehicle for the next two steps using the previous two-step speed and $v_{f}$ via linear extrapolation, $u_k^*$ and $v_{k+1}^*$ can be determined as follows:
\begin{align}
	& \bm{u}^* \!=\! \underset{(u,u^{+}) \in \bm{U}}{\mathop{\arg\min}}
	\!\! \left\{ C(u,v) \!+\! C(u^+,v^+) \!+\! V(d^{++},v^{++},v_{f}^{++}) \right\}  , \nonumber \\
	& v^{+*} \!=\! v + a(u^*) \Delta t, \quad v^{++*} \!=\! v^{+*} + a(u^{+*}) \Delta t \,,
\end{align}
where $\bm{u}=(u,u^{+})$ is a concatenation of two successive control inputs. After the forward computation, the actual implementation follows a receding horizon scheme, similar to model predictive control. In principle, this forward process can be extended to multistep look ahead, but the computational burden increases with the prediction horizon. To remedy this curse of dimensionality, one can rely on the ADP methods that are briefly introduced in Section~\ref{sec:method:ADP}.

\begin{figure}[t]
	\centering
	\includegraphics[width=1\linewidth]{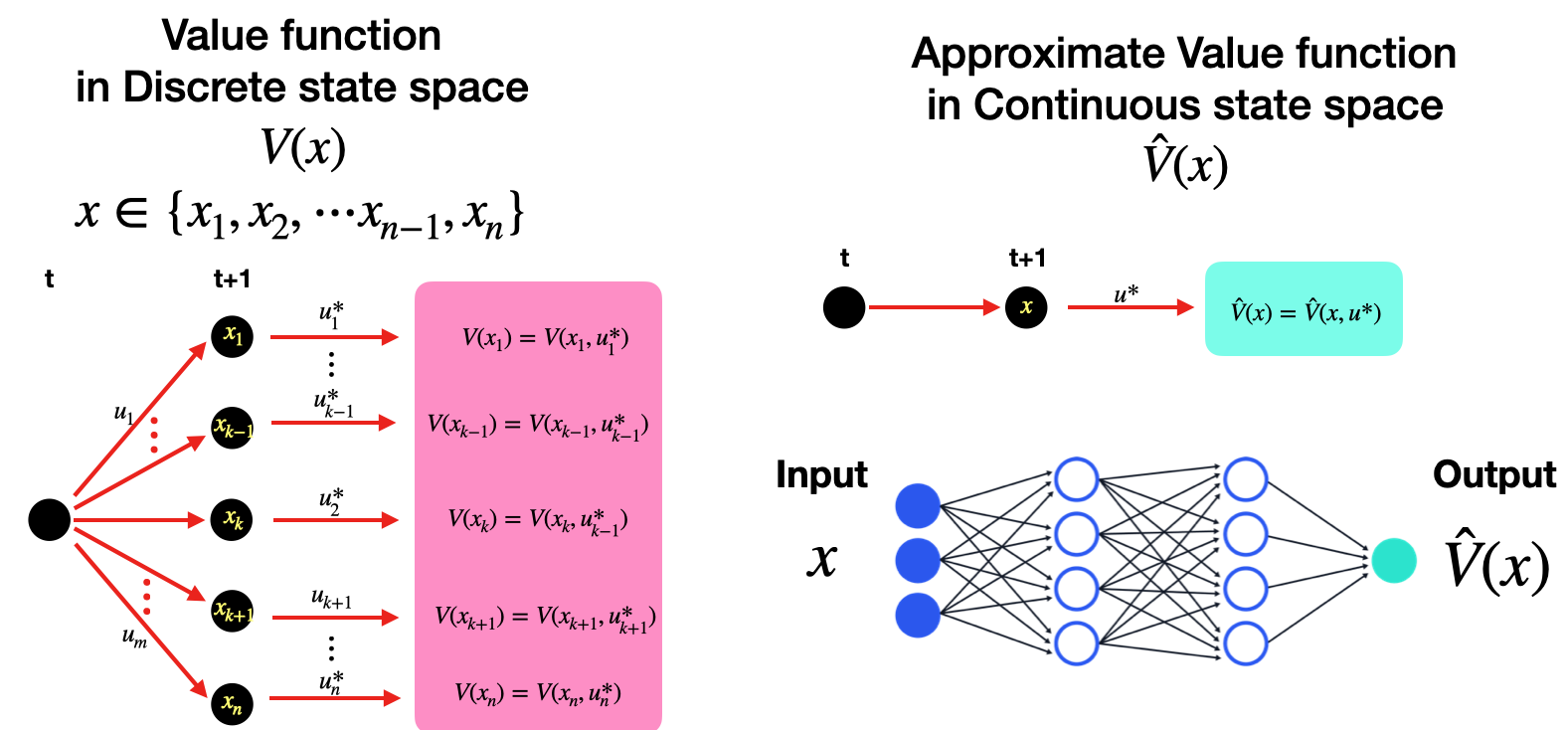}\vspace{-1mm}
	\caption{The value table obtained from DP-Backward was approximated with neural network as value function in continuous state space for application in ADP-method.}
	\label{fig:DPtoADP}
\end{figure}

\subsection{Approximate Dynamic Programming: Value Function Approximation}
\label{sec:method:ADP}
In the DP-based method, a value table was obtained for the set of quantized states $x=(d,v,v_f)$ in the DP-Backward process, and the energy-efficient optimal speed for the host vehicle was determined from the value table $V(d,v,v_f)$ in the DP-Forward process.
In the ADP-based method, the value table obtained in DP Backward~\ref{sec:method:DPB} was approximated as a function using an artificial neural network, as shown in Fig. ~\ref{fig:DPtoADP}. Neural networks (NNs) are universal approximators which are extensively used in machine learning and artificial intelligence. For this study of OCP corresponding to Eco-ACC, NNs approximating the associated value function are used for modeling the MPC-like forward optimization of the continuous state and inputs. More specifically, we used an NN having three layers with 70 units in each layer, and a sigmoid function was used as the activation function in the hidden layers. The error in the approximate value-function data from the DP-based value table was 0.0955 in the RRMSE.

\subsection{Reinforcement Learning}
\label{sec:method:RL}
\begin{figure}[t!]
	\centering
	\includegraphics[width=1\linewidth]{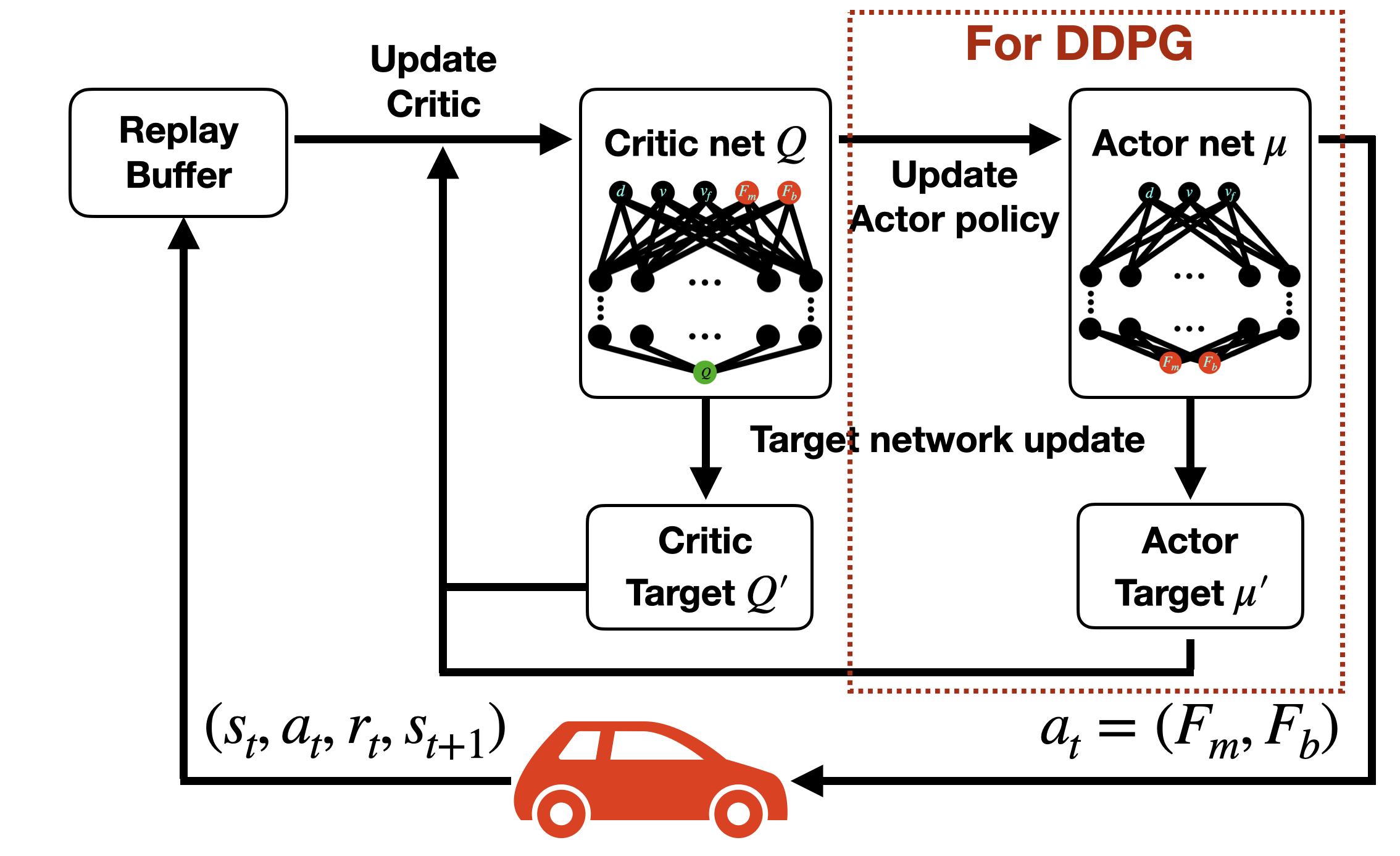}\vspace{-2mm}
	\caption{The structure of training RL agent process for Eco-ACC.}
	\label{fig:DQNDDPGacc}
\end{figure}

\begin{figure}[t!]
	\centering
	\includegraphics[width=1\linewidth]{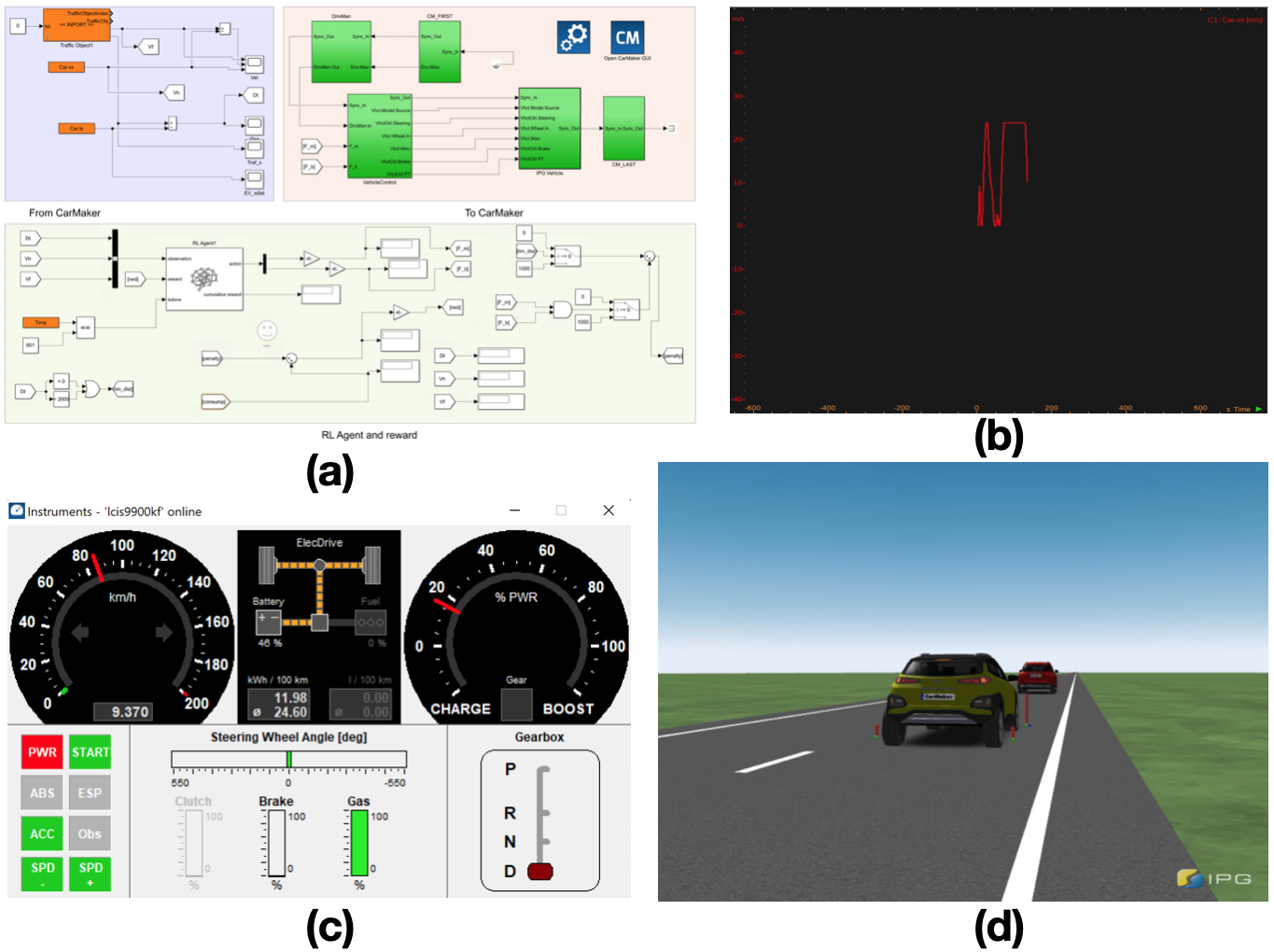}\vspace{-2mm}
	\caption{Training environment based on MATLAB/Simulink with connection to IPG CarMaker: (a) Simulink model with RL agent block, (b) Real-time vehicle speed, (c) CarMaker dashboard, and (d) CarMaker real-time simulation video.}
	\label{fig:rl_simulink}
\end{figure}

To allow the RL agents to learn the energy-efficient Eco-ACC policy from trial and error, the training environment shown in Fig.~\ref{fig:DQNDDPGacc} was established based on MATLAB/Simulink and IPG CarMaker, as shown in Fig.~\ref{fig:rl_simulink}.
During an episode, the simulation environment provides observations and reward based on the action. The transition $(s_t,a_t,r_t,s_{t+1})$ is stored in the replay buffer and is used to update the main critic $Q$ and main actor $\mu$ accordingly. And updates the target networks : $Q'$ and $\mu'$. At the next moment, the process of performing an action according to the deterministic policy and exploration noise is repeated.
For DQN, the critic networks $Q$ and $Q'$ are updated only and the control action is selected according to the $\epsilon$-greedy method. 

\begin{figure}[t!]
	\centering
	\includegraphics[width=1\linewidth]{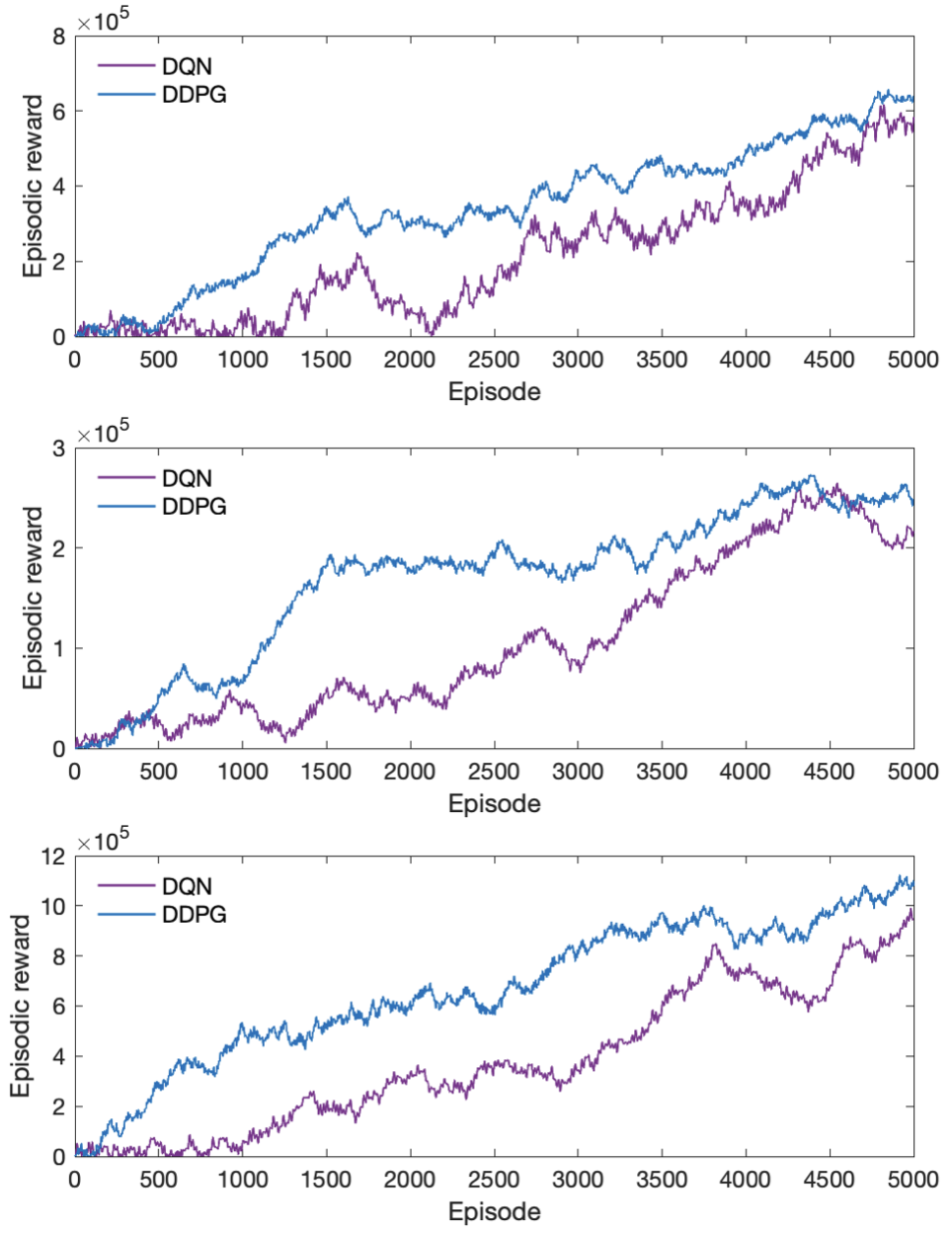}\vspace{-3mm}
	\caption{Episodic rewards in training process. In all three episodes, the agent initially received a small reward, but through a large number of learning, the reward received from the environment gradually increases.}
	\label{fig:RL_reward}
\end{figure}

In the training process, the reward received by the agent from the environment is shown in table~\ref{table:reward}. In order to minimize the energy consumption while driving on one episode, the negative value of motor power $-P_m$ .
In addition, always ensure that agents are rewarded by 1000, except when the distance to the vehicle ahead becomes 0 m (collision) or exceeds 2000 m (too much gap from the front vehicle). Finally, the agent is also rewarded, provided $F_m \cdot F_b \le 0$.
In this study, 5,000 training sessions were conducted for each scenario to train the RL agents. Fig.~\ref{fig:RL_reward} shows the episodic rewards computed during the  training process.

\begin{table}[t]
	\label{table:reward}
	\caption{Reward table for training DQN/DDPG agent.}\vspace{-1mm}
	\centering
	\begin{tabular}{|clllcc|}
	\hline
	\multicolumn{4}{|c|}{Objective or Condition}  & \multicolumn{2}{c|}{Value}  \\ \hline\hline
	\multicolumn{4}{|c|}{Minimizing energy consumption}  & \multicolumn{2}{c|}{$-P_m$} \\ \hline
	\multicolumn{4}{|c|}{\multirow{2}{*}
	{\begin{tabular}[c]{@{}c@{}}Maintaining distance gap range\\ $0<d<2000$\end{tabular}}} & True& 1000 \\ \cline{5-6} 
	\multicolumn{4}{|c|}{}  & False  & 0  \\ \hline
	\multicolumn{4}{|c|}{\multirow{2}{*}
	{\begin{tabular}[c]{@{}c@{}}Acceleration/Deceleration only\\ $F_m \cdot F_b \le 0$\end{tabular}}} & True & 1000 \\\cline{5-6} 
	\multicolumn{4}{|c|}{}   & False & 0 \\\hline
	\end{tabular}
\end{table}

\begin{table}[t]
	\label{table:hyperparam}
	\caption{Hyperparameters for training DQN/DDPG agent.}\vspace{-1mm}
	\centering
	\begin{tabular}{|cllccllc|}
		\hline
		\multicolumn{8}{|c|}{Hyperparameters}                                                                                                                                                                                                                                                                                \\ \hline
		\multicolumn{3}{|c}{Description}                                                                                         & \multicolumn{1}{c|}{Value}                & \multicolumn{3}{c}{Description}                                                                                         & Value               \\ \hline
		\multicolumn{3}{|c}{Actor network layers}                                                                                & \multicolumn{1}{c|}{2}                    & \multicolumn{3}{c}{Critic network layers}                                                                               & 3                   \\
		\multicolumn{3}{|c}{\multirow{2}{*}{\begin{tabular}[c]{@{}c@{}}Number of units per\\ critic network layer\end{tabular}}} & \multicolumn{1}{c|}{\multirow{2}{*}{256}} & \multicolumn{3}{c}{\multirow{2}{*}{\begin{tabular}[c]{@{}c@{}}Number of units per\\ critic network layer\end{tabular}}} & \multirow{2}{*}{70} \\
		\multicolumn{3}{|c}{}                                                                                                    & \multicolumn{1}{c|}{}                     & \multicolumn{3}{c}{}                                                                                                    &                     \\
		\multicolumn{3}{|c}{Actor learning rate}                                                                                    & \multicolumn{1}{c|}{0.01}                 & \multicolumn{3}{c}{Critic learning rate}                                                                                   & 0.01                \\ \hline
		\multicolumn{3}{|c}{Discount factor}                                                                                     & \multicolumn{1}{c|}{0.99}                 & \multicolumn{3}{c}{Replay buffer size}                                                                                  & 5000               \\
		\multicolumn{3}{|c}{Target smooth factor}                                                                                & \multicolumn{1}{c|}{0.005}                & \multicolumn{3}{c}{Batch size}                                                                                          & 64                  \\ \hline
	\end{tabular}
\end{table}

\begin{table}[t]
\label{table:vehicle}
\caption{Vehicle parameters for simulation.}\vspace{-1mm}
	\centering
	\begin{tabular}[b]{|@{\,\,}c@{\,\,}|@{\,\,}c@{\,\,}|@{\,\,}c@{\,\,}|}
		\hline
		Symbol & Description &      Value  \\
		\hline
		$m$ & \small{Vehicle mass}  &       \small{1600 [kg]} \\
		$C_d$ & \small{Coefficient of drag} &      \small{0.373 [-]} \\
		$A_f$ & \small{Frontal projected area of vehicle}  &     \small{2.0107 [$\mathrm{m^2}$]} \\
		$\rho_a$ & \small{Air density} &        \small{1.2  [$\mathrm{kg/m^3}$]} \\
		$C_r$ & \small{Coefficient of rolling resistance} &     \small{0.0088        [-]} \\
		$g$ & \small{Gravitational constant} &       \small{9.81  [$\mathrm{m^3/kg\cdot s}$]} \\
		$\gamma$ & \small{Gear ratio of final reduction drive} &        \small{7.4 [-]} \\
		$r_w$ & \small{Wheel radius} &      \small{0.326         [m]} \\
		$\alpha$ & \small{Road slope} &      \small{$0^\circ$} \\
		\hline
	\end{tabular}  
\end{table}

\section{Simulation Results}
\label{sec:sim}
To verify the strategy obtained from the method presented in Section~\ref{sec:method}, we considered three different scenarios according to the speed profile of the front vehicle. The front vehicle are assumed to drive according to the speed profiles of either
\begin{itemize}
\item
\emph{Scenario 1}: 
the HWFET cycle, representing highway driving conditions under 60 [mph],
\item
\emph{Scenario 2}:
the US06 cycle, representing high acceleration aggressive driving, or
\item
\emph{Scenario 3}:
the WLTP Class 3b cycle, representing European and Japanese commercial vehicles with a maximum speed of above 120 [kph].
\end{itemize}
The host vehicle follows the front vehicle while minimizing energy consumption. For DP and ADP, simulations were performed in IPG CarMaker~\cite{Carmaker} based on the speed profile of the host vehicle obtained from DP and ADP-Forward. For the RL method, a Simulink environment with a DDPG agent and reward function was constructed with the CarMaker extension. A video recording of the simulation is available at \smash{\url{https://youtu.be/DIXzJxMVig8}}.

\subsection{Dynamic Programming-Based method}
\begin{figure}[t]
	\centering
	\includegraphics[width=1\linewidth]{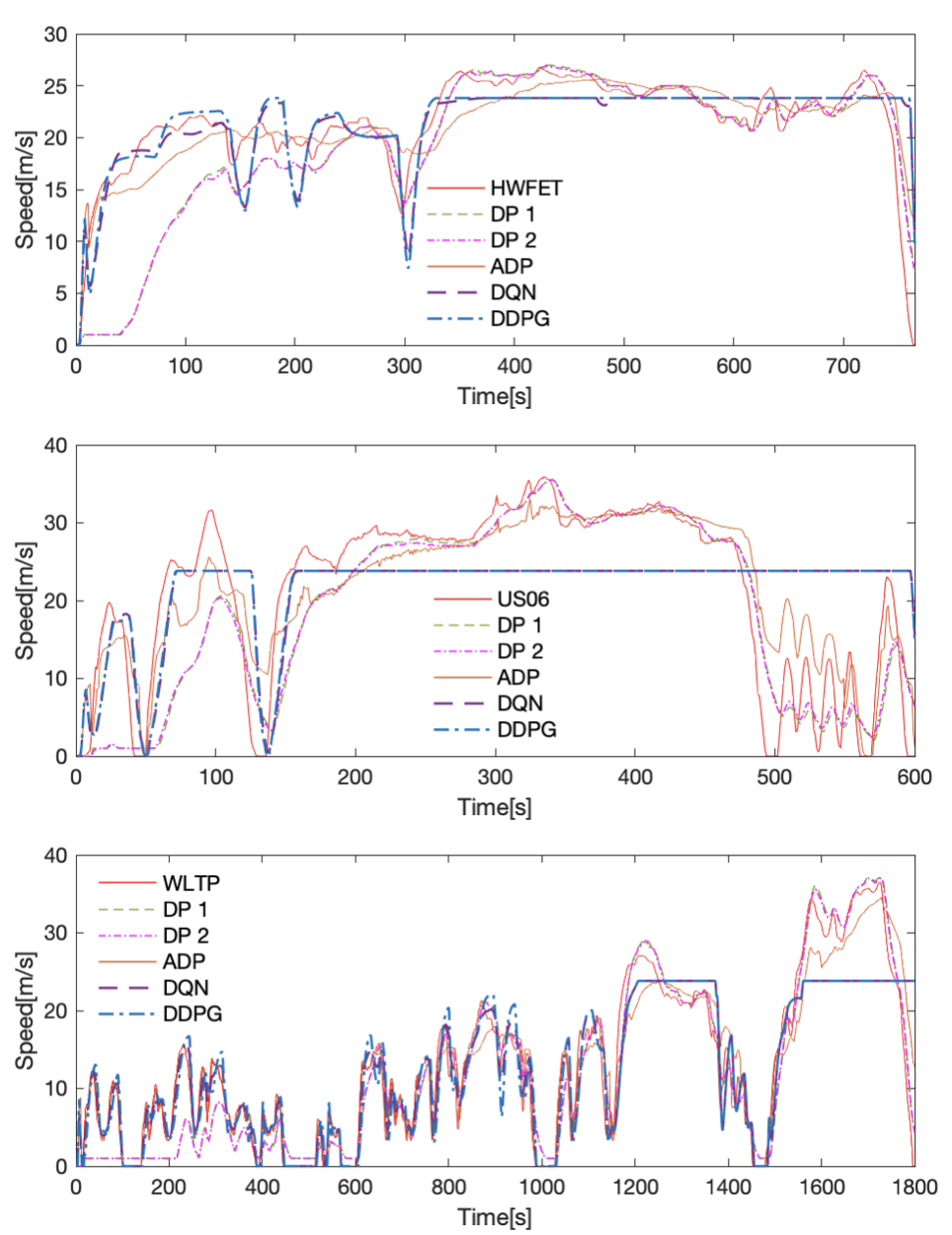}\vspace{-3mm}
	\caption{Comparisons of vehicle speed profiles for three driving scenarios of a front-vehicle: (1) the HWFET cycle (highway driving) and (2) the US06 cycle (city+highway driving) of the US  Environmental Protection Agency (EPA) federal test procedure (3) the WLTP Class 3b cycle.  DP1 refers to the one-step look ahead method in Section~\ref{sec:method:DPF1}, DP2 refers to the two-step look ahead method in Section~\ref{sec:method:DPF2}, ADP refers to the approximate dynamic programming method in Section~\ref{sec:method:ADP}, DQN and DDPG refers to the reinforcement learning method (DQN, DDPG) in Section~\ref{sec:method:RL}, respectively.}
	\label{fig:compare:v}
\end{figure}

\begin{figure}[t!]
	\centering
	\includegraphics[width=1\linewidth]{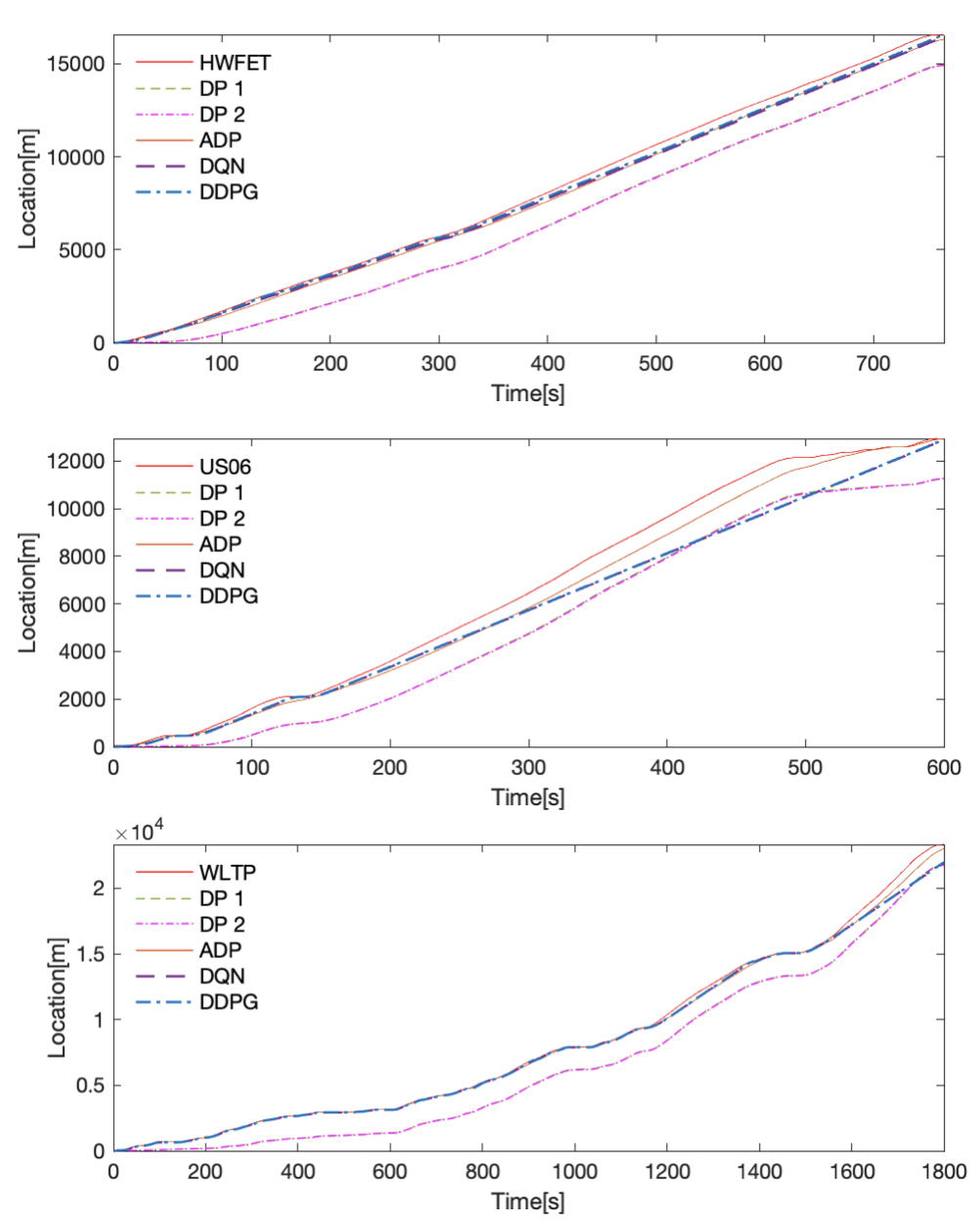}\vspace{-3mm}
	\caption{Comparison of vehicle travel distance for three driving scenarios of a front-vehicle.}
	\label{fig:compare:s}
\end{figure}

For the presented optimal control problem, a value table for each state was obtained from DP-Backward. Based on the value table, an optimal speed profile was obtained for the host vehicle. The simulation was conducted as follows. The virtual host vehicle was driven in CarMaker according to the generated speed profiles. The energy efficiency was measured using CarMaker.

It can be observed that the frequency of rapid acceleration and deceleration is lower than that of the front vehicle. In addition, a relatively gradual speed profile can be observed while maintaining the range of the distance gap. In other words, it can be observed that relatively gradual acceleration and deceleration strategies appear as a result of relatively high energy efficiency.

The same simulation procedure was performed for the front vehicle moving according to the US06 Profile.
Similar to Scenario 1, the speed profile obtained from the DP demonstrated better energy efficiency than that of the front vehicle with more moderate sensitivity/acceleration. Notably, it can be observed that the DP2 speed profile with 2-step lookahead indicates a relatively poor energy efficiency compared to DP1 with 1-step lookahead. This is because the distance gap with the front vehicle is initially widened, and then the speed of the front vehicle between 500 and 600 [s] is predicted at 2-step lookahead, which is a relatively rapid acceleration/deceleration compared to that of DP1 to ensure the maintenance of the range of the distance gap.

For the scenario with the front vehicle following the WLTP Class 3b driving cycle, both DP vehicles exhibited approximately 6\% higher energy efficiency than the front vehicle. It should be noted that, similar to Scenario 1, DP1 and DP2 showed almost the same energy efficiency. Because there is a relatively less rapid acceleration/deceleration of the front vehicle, the prediction of 2-step lookahead DP can affect the overall energy efficiency improvement, and the energy efficiency of the DP1 and DP2 results did not show much difference.

\subsection{Approximate Dynamic Programming-Based Method}
Similar to the previous DP-based method, the speed profile is obtained by applying 2-step lookahead and the energy efficiency is measured by IPG CarMaker.
The simulation results for Scenario 1 showed that the energy efficiency of ADP was 3.5\% higher than that of the front vehicle and 0.4\% higher than that of DP2. In the case of DP2, acceleration begins when the distance gap approaches the constraint limit of 2000 [m], whereas ADP starts acceleration when the distance gap is less than DP2. It can also be observed that the energy efficiency is similar even though it travels more distance than DP2. Similar to DP2, ADP exhibited better energy efficiency than the front vehicle. However, because acceleration begins relatively earlier than DP2, rapid acceleration and deceleration between 0 and 50 [s] is presumed to be the cause of the energy efficiency was lower by 6.3\% compared to DP2. For Scenario 3, ADP showed more travel distance and higher energy efficiency compared with the front vehicle and DP2.

\subsection{Reinforcement Learning-Based method}
The traveling speed and distance profiles shown in Figs. ~\ref{fig:compare:v} and~\ref{fig:compare:s}, respectively, were obtained using the 5,000-times trained DQN and DDPG agents. Unlike the DP and ADP results, the RL-host vehicle tends to drive while maintaining the maximum energy-efficient speed (approximately $23$ [m/s]). In particular, the US06 scenario achieved the result of driving at a constant speed despite the large distance from the vehicle ahead. However, the RL-Host vehicle does not drive fast in the WLTP scenario. The vehicle accelerates and decelerates in the low-speed range, and, in the latter part of the scenario, it tends to drive at a speed of approximately $23$ [m/s], except for the low-speed range of 1400$\sim$1600 [s]. As results, energy efficiency was improved by about 13.9\% (for DQN) and 14.8\% (for DDPG) on average of three scenarios when compared to the front vehicle.

\begin{table}
	\label{table:comparisons}
	\caption{Comparisons of energy efficiency.}
	\centering
	\begin{tabular}{|cccc|}
		\hline
		\multicolumn{4}{|c|}{Scenario 1 : HWFET front vehicle}                                                                                                                             \\ \hline
		\multicolumn{1}{|c|}{Car}                           & \multicolumn{1}{c|}{Travel Distance {[}m{]}}          & \multicolumn{1}{c|}{Efficiency {[}km/kWh{]}}        & Ratio {[}\%{]} \\ \hline
		\rowcolor[HTML]{FFE9E8} 
		\multicolumn{1}{|c|}{\cellcolor[HTML]{FFE9E8}HWFET} & \multicolumn{1}{c|}{\cellcolor[HTML]{FFE9E8}16556.08} & \multicolumn{1}{c|}{\cellcolor[HTML]{FFE9E8}8.5543} & 100.0          \\
		\multicolumn{1}{|c|}{DP 1}                          & \multicolumn{1}{c|}{14955.48}                         & \multicolumn{1}{c|}{8.7719}                         & 102.5          \\
		\multicolumn{1}{|c|}{DP 2}                          & \multicolumn{1}{c|}{14922.23}                         & \multicolumn{1}{c|}{8.7873}                         & 102.7          \\
		\multicolumn{1}{|c|}{ADP}                          & \multicolumn{1}{c|}{16314.00}                         & \multicolumn{1}{c|}{8.8183}                         & 103.1          \\
		\rowcolor[HTML]{F9F8C7} 
		\multicolumn{1}{|c|}{\cellcolor[HTML]{F9F8C7}DQN}   & \multicolumn{1}{c|}{\cellcolor[HTML]{F9F8C7}16314.00} & \multicolumn{1}{c|}{\cellcolor[HTML]{F9F8C7}8.8183} & 104.3          \\
		\rowcolor[HTML]{DBFFFE} 
		\multicolumn{1}{|c|}{\cellcolor[HTML]{DBFFFE}DDPG}    & \multicolumn{1}{c|}{\cellcolor[HTML]{DBFFFE}16222.86} & \multicolumn{1}{c|}{\cellcolor[HTML]{DBFFFE}8.9686} & 104.8          \\
		\hline
		\hline
		\multicolumn{4}{|c|}{Scenario 2 : US06 front vehicle}                                                                                                                        \\ \hline
		\multicolumn{1}{|c|}{Car}                          & \multicolumn{1}{c|}{Travel Distance [m]}              & \multicolumn{1}{c|}{Efficiency [km/kWh]}            & Ratio {[}\%{]} \\ \hline
		\rowcolor[HTML]{FFE9E8} 
		\multicolumn{1}{|c|}{\cellcolor[HTML]{FFE9E8}US06} & \multicolumn{1}{c|}{\cellcolor[HTML]{FFE9E8}12938.56} & \multicolumn{1}{c|}{\cellcolor[HTML]{FFE9E8}5.6370} & 100.0     \\
		\multicolumn{1}{|c|}{DP 1}                         & \multicolumn{1}{c|}{11280.08}                         & \multicolumn{1}{c|}{6.7159}                         & 119.1     \\
		\multicolumn{1}{|c|}{DP 2}                         & \multicolumn{1}{c|}{11283.19}                         & \multicolumn{1}{c|}{6.7431}                         & 119.6     \\
		\multicolumn{1}{|c|}{ADP}                          & \multicolumn{1}{c|}{12934.91}                         & \multicolumn{1}{c|}{6.3857}                         & 113.3     \\
		\rowcolor[HTML]{F9F8C7} 
		\multicolumn{1}{|c|}{\cellcolor[HTML]{F9F8C7}DQN} & \multicolumn{1}{c|}{\cellcolor[HTML]{F9F8C7}12689.71} & \multicolumn{1}{c|}{\cellcolor[HTML]{F9F8C7}6.7431} & 124.3     \\
		\rowcolor[HTML]{DBFFFE} 
		\multicolumn{1}{|c|}{\cellcolor[HTML]{DBFFFE}DDPG}   & \multicolumn{1}{c|}{\cellcolor[HTML]{DBFFFE}12704.54} & \multicolumn{1}{c|}{\cellcolor[HTML]{DBFFFE}7.0274} & 124.7     \\
		\hline
		\hline
		\multicolumn{4}{|c|}{Scenario 3 : WLTP class 3b front vehicle}                                                                                                               \\ \hline
		\multicolumn{1}{|c|}{Car}                          & \multicolumn{1}{c|}{Travel Distance [m]}              & \multicolumn{1}{c|}{Efficiency [km/kWh]}            & Ratio {[}\%{]} \\ \hline
		\rowcolor[HTML]{FFE9E8} 
		\multicolumn{1}{|c|}{\cellcolor[HTML]{FFE9E8}WLTP} & \multicolumn{1}{c|}{\cellcolor[HTML]{FFE9E8}23312.39} & \multicolumn{1}{c|}{\cellcolor[HTML]{FFE9E8}7.5131} & 100.0     \\
		\multicolumn{1}{|c|}{DP 1}                         & \multicolumn{1}{c|}{21807.45}                         & \multicolumn{1}{c|}{7.9681}                         & 106.1     \\
		\multicolumn{1}{|c|}{DP 2}                         & \multicolumn{1}{c|}{21763.19}                         & \multicolumn{1}{c|}{7.9681}                         & 106.1     \\
		\multicolumn{1}{|c|}{ADP}                         & \multicolumn{1}{c|}{23008.76}                         & \multicolumn{1}{c|}{8.3195}                         & 110.7     \\
		\rowcolor[HTML]{F9F8C7} 
		\multicolumn{1}{|c|}{\cellcolor[HTML]{F9F8C7}DQN}  & \multicolumn{1}{c|}{\cellcolor[HTML]{F9F8C7}21988.91} & \multicolumn{1}{c|}{\cellcolor[HTML]{F9F8C7}8.5071} & 113.2     \\
		\rowcolor[HTML]{DBFFFE} 
		\multicolumn{1}{|c|}{\cellcolor[HTML]{DBFFFE}DDPG}   & \multicolumn{1}{c|}{\cellcolor[HTML]{DBFFFE}22026.62} & \multicolumn{1}{c|}{\cellcolor[HTML]{DBFFFE}8.5616} & 113.9     \\ \hline
	\end{tabular}
\end{table}

\section{Conclusions and Future Work}
\label{sec:conclusion}
This paper proposed and compared learning-based methods (ADP, DQN and DDPG) for energy-efficient Eco-ACC for connected EVs.
An optimal control problem with constraints was formulated based on the EV model to minimize the energy consumption while satisfying the physical limits of actuation commands and avoiding a collision. For model-based learning, ADP, a neural network, was used for value function approximation, and multi-step look-ahead receding horizon control was applied in the forward direction. For model-free learning, we applied DQN and DDPG in which neural networks were used to train the state-action value function and deterministic policy. High-fidelity simulations using IPG CarMaker demonstrated the performance and effectiveness of the proposed learning methods for Eco-ACC.  
Our future work will involve performing hardware-in-the-loop-simulation (HiLS) using dSPACE MicroAutoBox II to further validate the feasibility of a real-time energy-efficient acceleration and deceleration strategy.

\bibliographystyle{IEEEtran}
\bibliography{ecoacc_RL}

\balance

\end{document}